\documentclass[aps,prb,twocolumn,superscriptaddress,10pt]{revtex4-2}
\usepackage[utf8]{inputenc}
\usepackage{libertine}
\usepackage[libertine]{newtxmath}
\usepackage{mathtools,graphicx,microtype,bm}
\usepackage[cal=boondoxo,frak=boondox]{mathalfa}
\usepackage[colorlinks=true,allcolors=blue]{hyperref} 

\begin{document}
\title{Quasi-bound States and Resonant Skew Scattering in Two-Dimensional Materials with a Mexican-Hat Dispersion}

\author{Vladimir A.\ Sablikov}
\email[E-mail:]{sablikov@gmail.com} 
\author{Aleksei A.\ Sukhanov}
\email[E-mail:]{aasukhanov@yandex.ru}
\affiliation{Kotelnikov Institute of Radio Engineering and Electronics, Fryazino Branch, Russian Academy of Sciences, Fryazino, Moscow District, 141190, Russia}

\begin{abstract}
Mexican-hat dispersion of band  electrons in two-dimensional materials attracts a lot of interest, mainly due to the Van Hove singularity of the density of states near the band edge. In this paper, we show that there is one more feature of such a dispersion, which also leads to nontrivial effects. It consists in the fact that the sign of the effective mass in the momentum space near the central extremum is opposite to the sign of the mass outside this region. For this reason, any localized potential that repels quasiparticles in the outer region attracts quasiparticles in the central region and thereby creates quasi-bound states. We study these states in the case when the Mexican-hat dispersion is formed due to the hybridization of the inverted electron and hole bands, and the potential is created by a point defect. The energy and width of the resonance of the local density of states corresponding to a quasi-bound state are found, and it is shown that, under certain conditions, a quasi-bound state can transform into a bound state in a continuum of band states. The presence of quasi-bound states leads to nontrivial effects in the spin-dependent scattering of electrons. Due to the quasi-bound state, the skew scattering is strongly enhanced for electrons with energy near the resonance, and the skewness angle varies over a wide range depending on the energy. In addition, in a certain energy range, a nontrivial effect of scattering suppression appears in the direction opposite to the skewness angle.
\end{abstract}

\maketitle

\section{Introduction}
Mexican-hat shaped dispersion of electronic spectrum is a relatively common property of many two-dimensional (2D) materials such as topological insulators in which the Mexican-hat dispersion naturally appears as a consequence of the band inversion~\cite{PhysRevB.85.161101}, bilayer graphene~\cite{mccann2007low,doi:10.1126/sciadv.aau0059}, monolayers of group III-IV chalcogenides~\cite{PhysRevB.95.115409,doi:10.1063/1.4928559,C5NR08982D}. The main interest to the Mexican-hat dispersion is usually attracted due to a Van Hove singularity of density of states close to the band edge, which opens the way for many striking effects caused by electron correlations, including the formation of a stable ferromagnetic phase~\cite{PhysRevB.75.115425,PhysRevLett.114.236602}, stimulation of electron pairing~\cite{PhysRevLett.56.2732,PhysRevLett.98.167002}, and dramatic changes in the spectrum of the bound state in the attractive potential~\cite{PhysRevLett.96.126402,PhysRevB.89.041405}. Recent experiments reveal sharp peaks in optical conductivity due to this feature of the density of states~\cite{PhysRevB.101.121115}. 

In this paper, we show that there is another feature of the Mexican-hat dispersion which also leads to nontrivial effects. This feature consists in the fact that the sign of the effective mass in the momentum space near the central extremum is opposite to the sign of the mass outside this region. For definiteness, consider a Mexican-hat dispersion in the conduction band. The effective mass of the electrons with the wave vectors near the central maximum is negative, while outside this region the effective mass is positive. Therefore, it can be expected that a negatively charged defect that normally repels band electrons will attract electrons with momenta near the center of the Mexican hat and thereby create a quasi-bound state against the background of a continuum of band states. We show that these states do exist and study their properties in the case when the Mexican-hat dispersion is due to the hybridization of electron-like and hole-like band states described within the frame of the Bernevig--Hughes--Zhang model~\cite{BHZ}. 

The energy of quasi-bound states lies above the central maximum of the Mexican hat, where they form resonances of the local density of states. The key role in the mechanism of the formation of quasi-bound states is played by the hybridization of the electron and hole bands. In particular, the hybridization parameter largely determines the width of the resonance. But besides this, the resonance width also depends on some overlap integral of the localized component of the wave function and the wave function of the continuum states near the defect. Because of this, the quasi-bound states have an interesting feature: under certain conditions, the resonance width can vanish and the quasi-bound state turns into a bound state in a continuum.

An interesting question is about the possible manifestations of quasi-bound states in the experiment. To this end, let us study how the quasi-bound state manifests itself in the process of scattering of band electrons. It turns out that the quasi-bound state strongly enhances the spin-dependent skew scattering of electrons with energies close to resonance ones, and as the energy changes, the skewness angle changes from 0 to $\pi$. In addition, in a certain energy range, a nontrivial effect of spin-dependent scattering suppression occurs in the direction opposite to the skewness angle for each spin.

\section{Quasi-bound states in a repulsive potential}\label{S:quasi-bound_state}
As a model of Mexican-hat dispersion we use four-band model of Bernevig, Hughes, and Zhang (BHZ) in which such a dispersion arises due to inversion and hybridization of the electron- and hole-like bands. For simplicity the model is supposed to be symmetric with respect to the electron and hole bands. The hybridization is conveniently described by a dimensionless parameter $a=A(B M)^{-1/2}$, where $A$, $B$, and $M$ are standard parameters of the BHZ model~\cite{BHZ}. $M$ is the mass term, $B$ the parameter describing the electron and hole band dispersion, and $A$ is the dimensional hybridization parameter. A repulsive nonmagnetic defect is described by a potential $V(r)$ which is supposed to be axially symmetric. 

The system Hamiltonian splits into two 2$\times$2 matrixes for spin-up and spin-down states. To be specific, in what follows we will consider the spin-up sector. In the dimensionless form the spin-up Hamiltonian reads
\begin{equation}
    H_{\uparrow}= 
    \begin{pmatrix}
        -1+\mathbf{\hat{k}}^2 & a\hat{k}_{+}\\
       a\hat{k}_{-} & 1-\mathbf{\hat{k}}^2\\
    \end{pmatrix}
    + v(r) \cdot \mathrm{I}_{2\times 2},
\label{h0}    
\end{equation}
where the values of the energy dimension are normalized to $|M|$. The distance is normalized to $\sqrt{|B/M|}$, the wave vector $k$ is normalized to $\sqrt{|M/B|}$, and $k_{\pm}=k_{x}\pm i k_{y}$. The normalized potential is denoted by $v(r)$. The spinor wave function $\Psi=(\psi_1, \psi_2)^T$ contains two components one of which, $\psi_1$, describes the contribution of the electron-like orbital and the other, $\psi_2$, the hole-like one. 

The Mexican-hat dispersion arises when $a<\sqrt{2}$, and the Mexican-hat shape is the more pronounced the smaller $a$. Therefore it is interesting to study the situation where $a\ll 1$. The energy of the Mexican hat bottom is $E_{min}=|a|\sqrt{1-a^2/4}$, and the central extremum is $E_{top}=1$. 

Quantum states induced by a defect potential are not difficult to study analytically when $a^2\ll 1$. In the limiting case of $a=0$ the Hamiltonian~(\ref{h0}) is diagonal, and the Schr\"odinger equation $H_{\uparrow}\Psi=E\Psi$ splits into two separate equations for the components of the spinor wave function $\Psi(\mathbf{r})=(\psi(\mathbf{r}), \phi(\mathbf{r}))^T$ which are easily analyzed. This opens up the possibility to study the Mexican-hat situation by considering $a$ as a small parameter.

In such a way we present the Hamiltonian~(\ref{h0}) as the sum of the Hamiltonians of two subsystems, $H_c$ and $H_b$, weakly coupled to each other via the Hamiltonian $W$
\begin{equation}
  H_{\uparrow}=H_c+H_b+W\,,  
\label{F-A_hamiltonian}
\end{equation}
where
\begin{equation}
  H_c=
  \begin{pmatrix}
    \mathbf{\hat{k}}^2-1 + v(r) & 0\\
   0 & 0\\
\end{pmatrix},\;
H_b=
\begin{pmatrix}
  0 & 0\\
 0 & -\mathbf{\hat{k}}^2+1 + v(r)\\
\end{pmatrix},
\end{equation}
and 
\begin{equation}
W = a
\begin{pmatrix}
    0 & \hat{k}_{+}\\
   \hat{k}_{-} & 0\\
  \end{pmatrix}.
\end{equation}
The Hamiltonians $H_c$ and $H_b$ describe uncoupled electron-like and hole-like subsystems.

The electron-like subsystem is described by the spinor $\Psi^{(c)}_{\mathbf{k}}=(\psi_\mathbf{k}, 0)^T$, where $\psi_\mathbf{k}(\mathbf{r})$ is defined by the equation
\begin{equation}
    [\hat{k}^2 + v(r)]\psi_{\mathbf{k}} = (1+E)\psi_{\mathbf{k}}\,.
\end{equation}
It is clear that $\Psi^{(c)}_{\mathbf{k}}$ is a continuum of states generated by the electron orbitals $|E\uparrow\rangle$ with the energy $E>-1$. Because of the axial symmetry, $\Psi^{(c)}_{\mathbf{k}}$ can be presented as a sum of angular harmonics: $\Psi^{(c)}_{\mathbf{k}}=\sum_m \Psi^{(c)}_{k,m} e^{i m \varphi}$, where $\Psi^{(c)}_{k,m}=(\psi_{k,m},0)^T$ and $k=\sqrt{1+E}$ plays the role of the wave number defined at $r\to \infty$. Asymptotically, as $r\to \infty$, the function $\psi_{k,m}$ behaves like $\sim C_1 J_m(kr)+C_2 Y_m(kr)$. Close to the center, the amplitude of $\psi_{k,m}$ is suppressed by the repulsive potential $v(r)$.

The Hamiltonian $H_b$ describes the states $\Phi=(0, \phi)^T$ formed by the hole orbitals $|H\uparrow\rangle$, where $\phi(\mathbf{r})$ is a solution of the equation
\begin{equation}
    [\hat{k}^2 - v(r)]\phi = (1-E)\phi\,.
    \label{e:bs}
\end{equation}
This equation has both discrete and continuous spectrum of eigenstates. In the energy range $E<1$, the spectrum is continuous, $E(q)$, and depends on one quantum number $q=\sqrt{1-E}$, which can be considered as a wave vector defined only at $r\to \infty$. The discrete spectrum lies at $E>1$. The discrete levels $E_{n,m}$ are determined by radial and angular quantum numbers $n$ and $m$, and are written as $E_{n,m}=1+\varepsilon_{n,m}$. The bound-state wave function has the form $\phi_{n,m}(r)e^{im\varphi}$.

\begin{figure}
    \centerline{\includegraphics[width=0.75\linewidth]{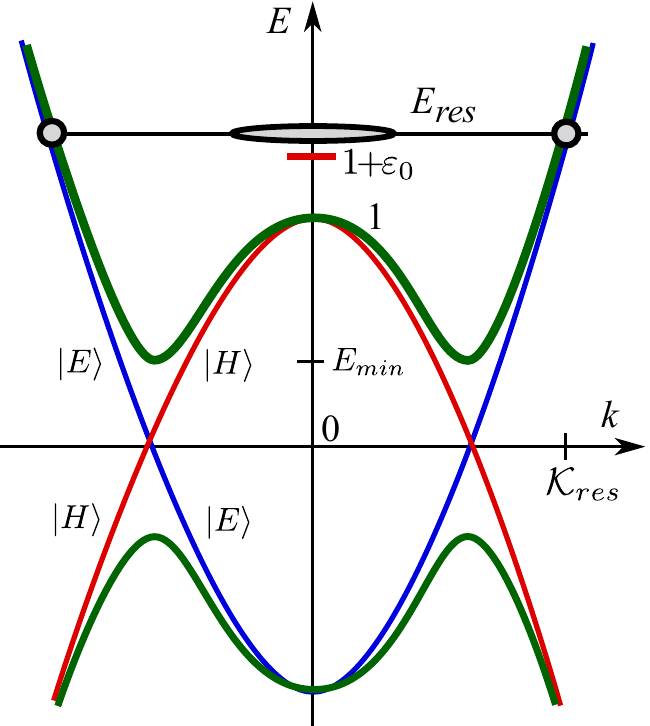}}
 	\caption{Schematic of the quasi-bound state formation due to coupling subsystems described by $H_b$ and $H_c$. Thin blue and red lines are the spectra of the electron-like and hole-like states, with $\varepsilon_0$ being a bound state in the hole subsystem. The thick green lines represent the band dispersion resulting from hybridization. The elongated ellipse symbolically depicts a quasi-bound state with resonant energy $E_{res}$.}
    \label{fig1}
\end{figure}

The Hamiltonian $W$ in Eq.~(\ref{F-A_hamiltonian}) couples the electron and hole subsystems. The situation is qualitatively illustrated in Fig.~\ref{fig1}. The coupling of the subsystems via $W$ leads to the formation of the resonant states and avoided crossing of the electron- and hole-like bands.

Thus, in the energy region $E>1$, the situation is similar to the Fano--Anderson problem of a localized state in a continuum. For simplicity, we assume here that the energy distance between the discrete energy levels is large enough and restrict ourselves to considering only one localized state with energy $1+\varepsilon_0$ and zero angular momentum. Following to the method of the Fano--Anderson theory~\cite{PhysRev.124.1866,MahajanJPhysA2006}, the wave function of the total system in the energy region $E>1$ can be constructed in the form
\begin{equation}
   \Psi=\sum\limits_{k,m} A_{k,m} \Psi^{(c)}_{k,m}e^{im\varphi} + B\,\Phi\,.
   \label{e:F-A_func1}
\end{equation}
Here we neglected also the $\Phi_{q,m}$ states of the continuous part of the spectrum of the hole subsystem, since they are far  from the considered state in energy, as are the excited states of the discrete spectrum. Substituting the wave function (\ref{e:F-A_func1}) into the Schr\"odinger equation we arrive at the following eigenfunctions of the Hamiltonian $H_{\uparrow}$:
\begin{align}
    \Psi_{\mathcal{K}}=B_{\mathcal{K}}\Bigl[\Phi & +a Z_{\mathcal{K}}\sum_{m} \langle \mathcal{K},m|\hat{k}_+|\Phi\rangle \Psi^{(c)}_{\mathcal{K},m}\Bigr.\nonumber \\
    \Bigl. & +a \sum\limits_{k,m} \mathcal{P}\frac{1}{\mathcal{K}^2-k^2}\langle k,m|\hat{k}_+|\Phi\rangle \,\Psi^{(c)}_{k,m} \Bigr]\,, 
 \end{align}
 where $\mathcal{P}$ denotes the principal value. The wave functions are labeled by the quantum number $\mathcal{K}$ which is related to the energy $E$ by $\mathcal{K}=\sqrt{E+1}$. The value $Z_{\mathcal{K}}$ is defined as
\begin{equation}
    Z_{\mathcal{K}}=\frac{\mathcal{K}^2-2-\varepsilon_0 -\Sigma_{\mathcal{K}}}{a^2\sum_m|\langle \mathcal{K},m|\hat{k}_+|\Phi\rangle|^2}\,,
\end{equation}
with $\Sigma_{\mathcal{K}}$ being the self-energy function
\begin{equation}
    \Sigma_{\mathcal{K}} = a^2\sum_{k,m} \mathcal{P}\frac{1}{\mathcal{K}^2-k^2}\langle k,m|\hat{k}_+|\Phi\rangle\,.
\end{equation}
The amplitude $B_{\mathcal{K}}$ is
\begin{equation}
    B_{\mathcal{K}}=\frac{1}{\sqrt{(\mathcal{K}^2-2-\varepsilon_0-\Sigma_{\mathcal{K}})^2+\gamma_{\mathcal{K}}^2}}\,,
\label{amplitude_res}
\end{equation}
where
\begin{equation}
    \gamma_{\mathcal{K}}=\frac{a^2L}{2v_{\mathcal{K}}}\sum_m |\langle \mathcal{K},m|\hat{k}_+|\Phi\rangle|^2\,,
\end{equation}
with $L$ being the normalization length and $v_{\mathcal{K}}=dE/d\mathcal{K}=2{\mathcal{K}}$.

Since the angular quantum number of the localized state $\Phi$ we are considering is zero, it is clear that the matrix element $\langle k,m|\hat{k}^+|\Phi\rangle$ in nonzero only for $m=1$, and in the above equations the sum over $m$ reduces to only one term with $m=1$. Therefore
\begin{align}
    \Sigma_{\mathcal{K}} & = a^2\sum_{k} \mathcal{P}\frac{1}{\mathcal{K}^2\!-\!k^2}\langle k,1|\hat{k}_+|\Phi\rangle,\label{res_self}\\ 
    \gamma_{\mathcal{K}} & = \frac{a^2L}{2v_{\mathcal{K}}}|\langle \mathcal{K},1|\hat{k}_+|\Phi\rangle|^2. \label{res_width}
\end{align}

Equation~(\ref{amplitude_res}) clearly shows that the amplitude of the wave functions has a resonance at the energy
\begin{equation}
    E_{res}=1+\varepsilon_0+\Sigma_{\mathcal{K}_{res}}\,,
\label{res_energy}
\end{equation}
which lies above the central maximum of the Mexican-hat profile, and the resonance width function is given by Eq.~(\ref{res_width}).

The resonance of the local density of states near the defect arises at any repulsive potential that can create a bound state according to Eq.~(\ref{e:bs}). The resonance width is determined by two factors: the hybridization of the electron and hole bands described by the parameter $a$, and the overlap of the localized and continuum wave functions which is determined by the potential $v(r)$. The dependence of the resonance width on the parameter $a$ is approximated as $\gamma_{\mathcal{K}}\propto a^2$ for $a^2\ll 1$. 

The dependence on $v(r)$ is more complicated and can lead to nontrivial consequences. The point is that the matrix element $\langle \mathcal{K}_{res},1|\hat{k}_+|\Phi\rangle$ can vanish under certain conditions regarding the function $v(r)$. In this case, the resonance width vanishes and the quasi-bound state transforms into a bound state in a continuum (BIC). The condition under which this occurs are determined by two equations. The first is $\langle \mathcal{K}_{res},1|\hat{k}_+|\Phi\rangle = 0$ and the second is Eq.~(\ref{res_energy}) that determines $\mathcal{K}_{res}$. Both equations contain $v(r)$. The BIC is formed if these equations are compatible and have a common root $\mathcal{K}_{res}$. 

We have explored this possibility for some specific forms of the potential. In the case of a step-like form $v(r)=v_0 \Theta(r_0-r)$, which models a short-range interaction, the above equations are compatible only for certain values of the pair of parameters $r_0$ and $v_0$. The BIC is formed only at these values of the amplitude and radius of the potential. In the case of the Coulomb potential $v(r)=Z/r$, the BIC cannot be formed from the ground-state $\phi_{0,0}$, however the first exited state $\phi_{1,0}$ can give rise to the BIC\@. This mechanism of BIC formation corresponds to the trend of recent years to consider resonant states as the basis for the formation of BIC and quasi-BIC resonances in various physical systems~\cite{hsu2016bound,PhysRevLett.123.253901}.

The results presented refer to spin-up states. For spin-down states, it is possible not to carry out a separate calculation, but to use a unitary transformation that relates the spin-up and spin-down Hamiltonians: $\hat{H}_{\downarrow} = \hat{U}\hat{H}_{\uparrow}\hat{U}^{-1}$ where $\hat{U}= K\tau_z$, $K$ denotes the complex conjugation operation and $\tau_z$ is Pauli matrix. The spin-down wave function is related to the spin-up one by $\Psi_{\downarrow}=\hat{U}\Psi_{\uparrow}$. The general property of the BHZ Hamiltonian is that the Kramers pair of the eigenstates corresponds to opposite spins moving in opposite directions, in this sense the states are helical.  

In this way we find that the spectrum of spin-down quasi-bound states is the same as that of spin-up states. But the wave functions change in accordance with their helical nature and in the case of a quasi-bound state correspond to rotation around the center in the opposite direction. For this reason, one can expect that they lead to nontrivial effects in spin-dependent scattering of electrons by defects with quasi-bound states.

\section{Electron scattering by defects with a resonant state}\label{S_scatter}

In this section we study the scattering of band electrons by a point-like defect with a repulsive potential that creates a quasi-bound state. Interest in this problem stems from our expectation that it is precisely in scattering that the quasi-bound states under study can manifest themselves most clearly.

The problem is solved on the basis of the Hamiltonian~(\ref{h0}) for spin-up states. According to the standard scattering theory~\cite{newton2013scattering}, the wave function should asymptotically, at $r\to \infty$, be the sum of an incident wave propagating along the $x$ axis and a scattered outgoing wave. For an ingoing particle in the conduction band with a wave vector $k$, this can be written as
\begin{equation}\label{e:scatter}
    \Psi = C_k 
    \begin{pmatrix}
      1 \\ -\beta_k  
    \end{pmatrix}
    e^{ikx}+ f(k,\varphi) \frac{e^{ikr}}{\sqrt{r}}\,,
\end{equation}
where $\beta_k=-ak/(\varepsilon_k-1+k^2)$, with $\varepsilon_k$ being the band dispersion.
The scattering amplitude $f(k,\varphi)$ is a spinor, which is  determined from the Schr\"odinger equation with the Hamiltonian~(\ref{h0}) and the boundary condition at infinity given by Eq.~(\ref{e:scatter}).

It is clear that, in the general case, the scattering amplitude depends on the shape of the potential, since it essentially determines the wave functions and the spectrum of quasi-bound states. For definiteness, we confine ourselves here to the simplest and most universal case of a potential of zero radius, when the wave function is predominantly outside the region of potential localization. However, even in this case there is a well-known problem of processing $\delta$-potential in 2D space, which requires some regularization procedure~\cite{Jackiw:1991delta, doi:10.1119/1.19485, geltman2011bound}. We overcome this problem by considering the zero-radius potential as the limit of a sequence of step potentials $v(r)=v_0\Theta(r_0-r)$ with decreasing radius $r_0\to 0$ and scaling the amplitude as $v_0 \sim r_0^{-2}$.

It is convenient to use the polar coordinates (radius $r$ and angle $\varphi$) and represent the wave function in the form
\begin{equation}
    \Psi(\mathbf{r})=\sum\limits_{m} 
    \begin{pmatrix}
       \psi_{1,m}(r)\\
       i \psi_{2,m}(r)e^{-i\varphi}
    \end{pmatrix}
    e^{im\varphi}\,,
\end{equation}
where the functions $\psi_{1,m}(r)$ and $\psi_{2,m}(r)$ are defined by the following equations which are obtained directly from Eq.~(\ref{h0}):
\begin{align}
     [E+1-\hat{k}_m^2-v(r)]\psi_{1,m} - a\left(\frac{d}{dr}-\frac{m-1}{r}\right)\psi_{2,m}&=0 \label{eq_polar1}\\
     a\left(\frac{d}{dr}+\frac{m}{r}\right)\psi_{1,m} + [E-1+\hat{k}_{m-1}^2-v(r)]\psi_{2,m}&=0\,, \label{eq_polar2}
\end{align}
where $\hat{k}_m^2=-\frac{1}{r}\frac{d}{dr}\left(r\frac{d}{dr}\right)+\frac{m^2}{r^2}$ and $E$ is the energy. 

In the case of the step-like potential, Eqs~(\ref{eq_polar1}) and (\ref{eq_polar2}) are solved exactly using the method described in detail in Ref.~\cite{PhysRevB.95.085417}. The solution is presented in terms of the Bessel functions. In each of the regions, $0<r<r_0$ and $r>r_0$, where the potential is constant, $v=v_0$ and $v=0$, the wave functions are constructed on two basis functions, which are selected from a set of Bessel functions $J_m, Y_m , H_m^{(1)}, H_m^{(2)}, I_m, K_m$ in accordance with the boundary conditions at $r\to 0$ and $r \to \infty$.

In the region $0<r<r_0$, the wave function is expressed in terms of the Bessel functions $J_m$ and $I_m$~\footnote{Here we have taken into account that for $v_0\gg 1$ and $E\gtrsim 1$ in the region $r<r_0$ there are two characteristic wave numbers $q$ and $i\lambda$, one of which is real and the other is imaginary.}
\begin{align}
\psi_{1,m}(r)=& A_{1,m} J_m(qr)+A_{2,m} I_m(\lambda r)\,, \label{psi1m_i} \\  
\psi_{2,m}(r)=& A_{1,m} \beta_{J,q} J_{m-1}(qr)+A_{2,m} \beta_{I} I_{m-1}(\lambda r)\,, \label{psi2m_i}
\end{align}
where 
\begin{align}
 q=&\sqrt{1-\frac{a^2}{2}+\sqrt{(E-v_0)^2-a^2\left(1-\frac{a^2}{4}\right)}}, \label{e:q}\\
 \lambda=&\sqrt{-1+\frac{a^2}{2}+\sqrt{(E-v_0)^2-a^2\left(1-\frac{a^2}{4}\right)}}, \label{e:lambda} 
\end{align}
and
\begin{equation}
    \beta_{J,q}=\frac{-aq}{E-v_0-1+q^2}, \quad \beta_{I}=\frac{-a\lambda}{E-v_0-1-\lambda^2}.
\end{equation}

In the outer region $r>r_0$, the wave function is constructed on the basis of three Bessel functions. Two functions are taken from the set $J_m$, $Y_m$, $H^{(1)}_m$, and $H^{(2)}_m$, and the third is the modified Bessel function of the second kind $K_m$. The asymptotic behavior corresponding to the scattering theory requirement given by Eq.~(\ref{e:scatter}) is provided by using the functions $J_m$, $H^{(1)}_m$, and $K_m$. 

Indeed, the incoming wave, having been expanded in the angular harmonics, is expressed in terms of the Bessel functions $J_m$: 
\begin{equation}\label{e:incident}
    \Psi_{in}=C_k \begin{pmatrix} 1\\-\beta_k \end{pmatrix} \sum\limits_m i^m J_m(kr)e^{im\varphi}\,. 
\end{equation}

The asymptotic behavior of the scattered wave is described by the Hankel functions $H_m^{(1)}(kr)$, and the modified Bessel functions $K_m(\kappa r)$ represent a component decreasing with increasing distance from the defect.

Thus, in the region $r>r_0$ the wave function has the form:
\begin{align}
    \psi_{1,m}(r)\!=& B_m J_m(kr)+C_m H_m^{(1)}(kr) + D_m K_m(\kappa r)\,, \label{psi1m_e} \\  
    \psi_{2,m}(r)\!=& B_m \beta_{J,k} J_{m-1}(kr)\!+\!C_m \beta_{J,k} H_{m-1}^{(1)}(kr)\!+\!D_m \beta_{K} K_{m-1}(\kappa r)\,, \label{psi2m_e}
\end{align}
where
\begin{align}
    k=&\sqrt{1-\frac{a^2}{2}+\sqrt{E^2-a^2\left(1-\frac{a^2}{4}\right)}}, \label{e:k}\\
    \kappa=&\sqrt{-1+\frac{a^2}{2}+\sqrt{E^2-a^2\left(1-\frac{a^2}{4}\right)}}, \label{e:kappa} 
   \end{align}
and
\begin{equation}
    \beta_{J,k}=\frac{-ak}{E-1+k^2}, \quad \beta_{K}=\frac{a\kappa}{E-1-\kappa^2}.
\end{equation}

According to Eq.~(\ref{e:incident}), the coefficient $B_m$ is $B_m=i^m C_k$. All other coefficients, $A_{1,m}, A_{2,m}, C_m$ and $D_m$, in Eqs.~(\ref{psi1m_i}), (\ref{psi2m_i}), (\ref{psi1m_e}) and (\ref{psi2m_e}) are determined from conditions for matching the wave functions and their derivatives at the boundary $r=r_0$.

In this way, we have calculated all the above coefficients as functions of the energy and the wave functions for a sequence of increasing values of $v_0\gg 1$, simultaneously decreasing the radius $r_0$ so that the product $v_0 r_0^2\equiv v$ remains a constant value until the scattering amplitude as a function of energy ceases to change qualitatively. The quantity $v/\pi$ is the amplitude of the equivalent $\delta$ potential. 

Of most interest are the coefficients $C_m$ that determine the amplitude of the scattered wave. For convenience, they will be normalized to corresponding coefficients $B_m$ so that $\widetilde{C}_m=C_m/B_m$. The most important are only $s$- and $p$-waves with the angular numbers $m=0$ and $m=1$, since the amplitudes of higher harmonics are small for a short-range potential, at least in the parameter $(kr_0)^{2m}$. This is quite analogous to the scattering theory for 2D systems with one and two band spectra~\cite{doi:10.1119/1.14623,PhysRevB.76.073411}. 

The calculation shows that the amplitude $\widetilde{C}_1(E)$ as a function of energy has a sharp peak located above the top of the Mexican-hat dispersion, which corresponds to the quasi-bound state. The width of the peak increases with the hybridization parameter $a$. Interestingly, the maximum of $\widetilde{C}_1(E)$ is preserved with increasing $a$ up to $a=\sqrt{2}$. As a function of the potential amplitude $v$, the value $\widetilde{C}_1$ gradually increases. 

\begin{figure}
    \centerline{\includegraphics[width=0.95\linewidth]{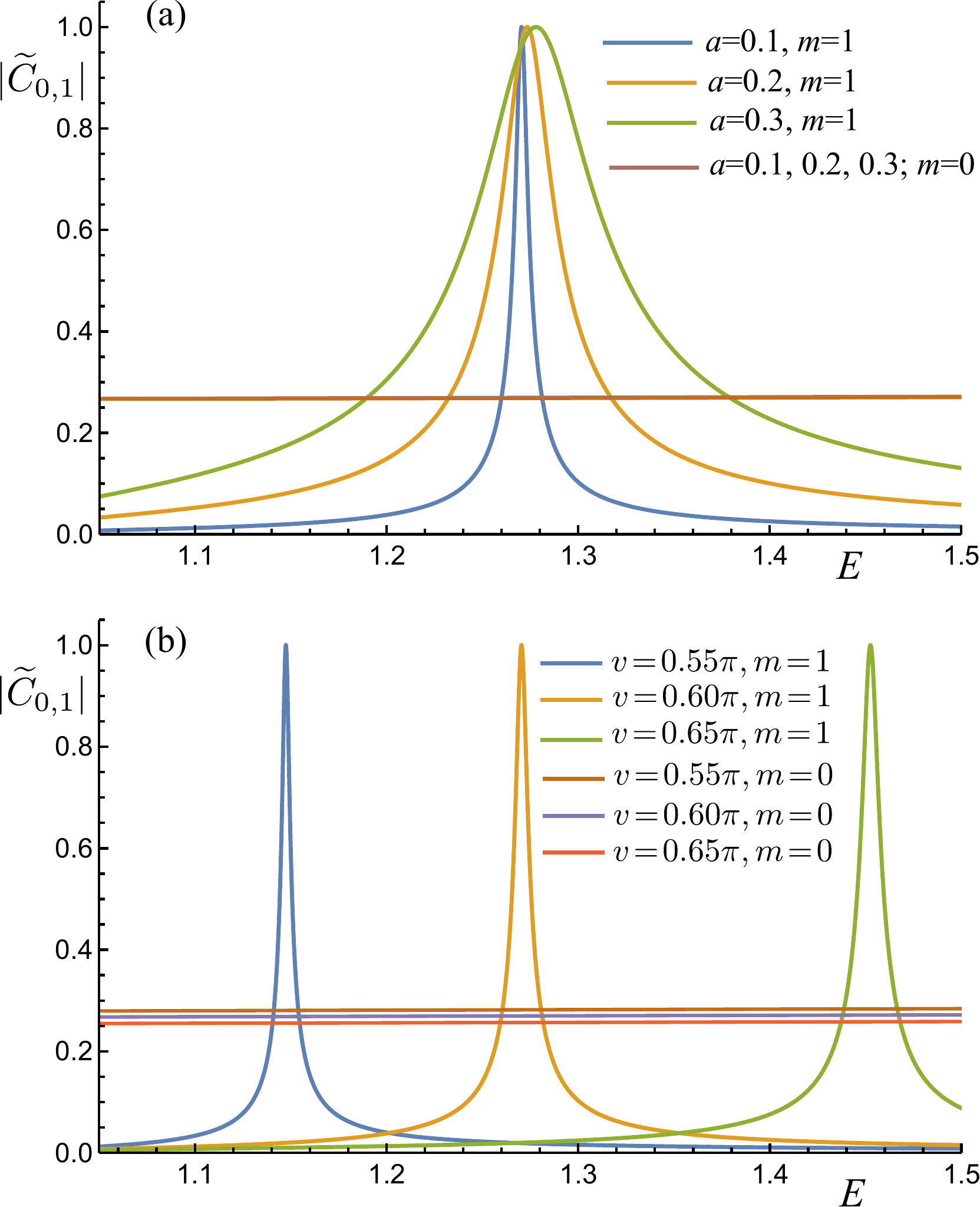}}
    \caption{The amplitudes of outgoing scattered $s$- and $p$-waves as functions of energy $E$ for different (a) hybridization parameters, $a$, and (b) the amplitudes, $v$, of the potential.}
    \label{fig2}
\end{figure}

The coefficient $\widetilde{C}_0(E)$ also has a resonant feature at a much higher energy, which, however, disappears with decreasing the radius $r_0$. In the vicinity of the $\widetilde{C}_1$ peak, the coefficient $\widetilde{C}_0$ is nearly constant. The energy dependence of $\widetilde{C}_1$ and $\widetilde{C}_0$ is shown in Fig.~\ref{fig2} for a variety of the hybridization parameter and the potential amplitude.

The coefficients $\widetilde{C}_1$ and $\widetilde{C}_0$ determine directly the scattering amplitude $f(r,\varphi)$. To find the scattering cross section, it is necessary to calculate the scattered current. The operator $\mathbf{j}$ is defined in the standard way via the velocity operator $\mathbf{v}=\frac{1}{\hbar}\frac{\delta H}{\delta \mathbf{k}}$ and the density operator. Having done these calculations for the BHZ Hamiltonian, we arrive at a general expression for the average current in a state described by the spinor $\Psi=(\psi_1, \psi_2)^T$, 
\begin{multline}
    \langle\Psi|\mathbf{j}|\Psi\rangle =-\psi_1\hat{\mathbf{k}}\psi_1^*+\psi_1^*\hat{\mathbf{k}}\psi_1+\psi_2\hat{\mathbf{k}}\psi_2^*-\psi_2^*\hat{\mathbf{k}}\psi_2\\ + a\left(\psi_1^*\psi_2 \mathbf{e}_+ +\psi_2^*\psi_1 \mathbf{e}_-\right)\,,
    \label{eq:current}
\end{multline}
where $\mathbf{e}_{\pm}$ is complex combination of the unit vectors: $\mathbf{e}_{\pm}=\mathbf{e}_x \pm i\mathbf{e}_y =e^{\pm i\varphi} \left(\mathbf{e}_{r}\pm i\mathbf{e}_{\varphi}\right)$ for the cartesian and polar coordinates, respectively. 

The differential cross section of the scattering is found as a ratio of the current density $j_{sc}$ scattered in the radial direction $\mathbf{e}_r$ to the incident flow $I_{in}$. Calculating $j_{sc}$ and $I_{in}$ as the average values of the current operator (\ref{eq:current}) over the scattered and incident wave functions we get
\begin{equation}
    \frac{d\sigma(E,\varphi)}{d\varphi}=\frac{2}{\pi k}\sum\limits_{m, m'}\widetilde{C}_m(E) \widetilde{C}_{m'}^*(E)e^{i(m-m')\varphi}\,,
    \label{eq:scatt_cross_section}
\end{equation} 
where $k=k(E)$ is defined by Eq.~(\ref{e:k}). Taking into account that for a short-range potential we can restrict ourselves to the angular harmonics $m=1$ and $m=0$, the expression for the cross section can be written as 
\begin{equation}
    \frac{d\sigma(E,\varphi)}{d\varphi}=\frac{2}{\pi k} S(E,\varphi)\,, 
    \label{eq:scatt_cross_section1}
\end{equation} 
where $S(E,\varphi)$ is a function that largely determines the energy and angle dependence of the scattering,
\begin{equation}
    S(E,\varphi)=|\widetilde{C}_1|^2+|\widetilde{C}_0|^2+2|\widetilde{C}_1\widetilde{C}_0|\cos \left(\varphi-\Delta\chi \right) \,.
    \label{e:S}
\end{equation}
Here $\widetilde{C}_0(E)$ and $\widetilde{C}_1(E)$ are represented as
\begin{equation}
    \widetilde{C}_{0,1}(E)=|\widetilde{C}_{0,1}(E)| e^{i\chi_{0,1}(E)},
\end{equation}
and
\begin{equation}\label{eq:d_chi}
    \Delta \chi(E)=\chi_0(E)-\chi_1(E).
\end{equation} 

The scattering cross section for spin-down electrons is found directly using the unitary transformation $\hat{U}$ introduced above. Thus, we come to a fairly obvious conclusion that for electrons with spin down $S_{\downarrow}(E,\varphi)=S(E,-\varphi)$ and the scattering cross section differs from Eq.~(\ref{eq:scatt_cross_section1}) only by changing the sign of $\varphi$.

It can be seen that the scattering cross section as a function of the energy and the scattering angle is mainly determined by the amplitudes and phases of $\widetilde{C}_0(E)$ and $\widetilde{C}_1(E)$. An analysis of Eq.~(\ref{eq:scatt_cross_section1}) shows that there are the following non-trivial features of the scattering:\\ 
\indent (i) The scattering cross section has a sharp resonance near the energy of the quasi-bound state.\\ 
\indent (ii) $d\sigma/d\varphi$ for each spin component is asymmetric with respect to the scattering angle, so that electrons with opposite spins scatter predominantly in opposite directions. The magnitude of this asymmetry near the resonance is quite large. The skewness angle, defined from the maximum of $d\sigma/d\varphi$ as a function of $\varphi$, is equal to $\pm \Delta \chi(E)$ for spin-up and spin-down electrons respectively.\\
\indent (iii) The skewness angle $\Delta \chi(E)$ changes in a wide range from 0 to $\pi$ depending on the energy when $E$ passes through the resonance. At the peak of the resonance in energy, the cross section as a function of angle reaches its maximum at $\varphi \approx \pi/2$, and on the slopes of the resonance, forward scattering (on the high-energy slope) or backward scattering (on the low-energy slope) become predominant.\\
\indent (iv) At a certain energy, defined by the equation $|\widetilde{C}_1(E)|=|\widetilde{C}_0(E)|$, the scattering cross section vanishes in the direction opposite to the skewness angle for each spin, $\varphi=\pm (\Delta \chi(E) + \pi)$.
\begin{figure}[h]
    \centerline{\includegraphics[width=0.95 \linewidth]{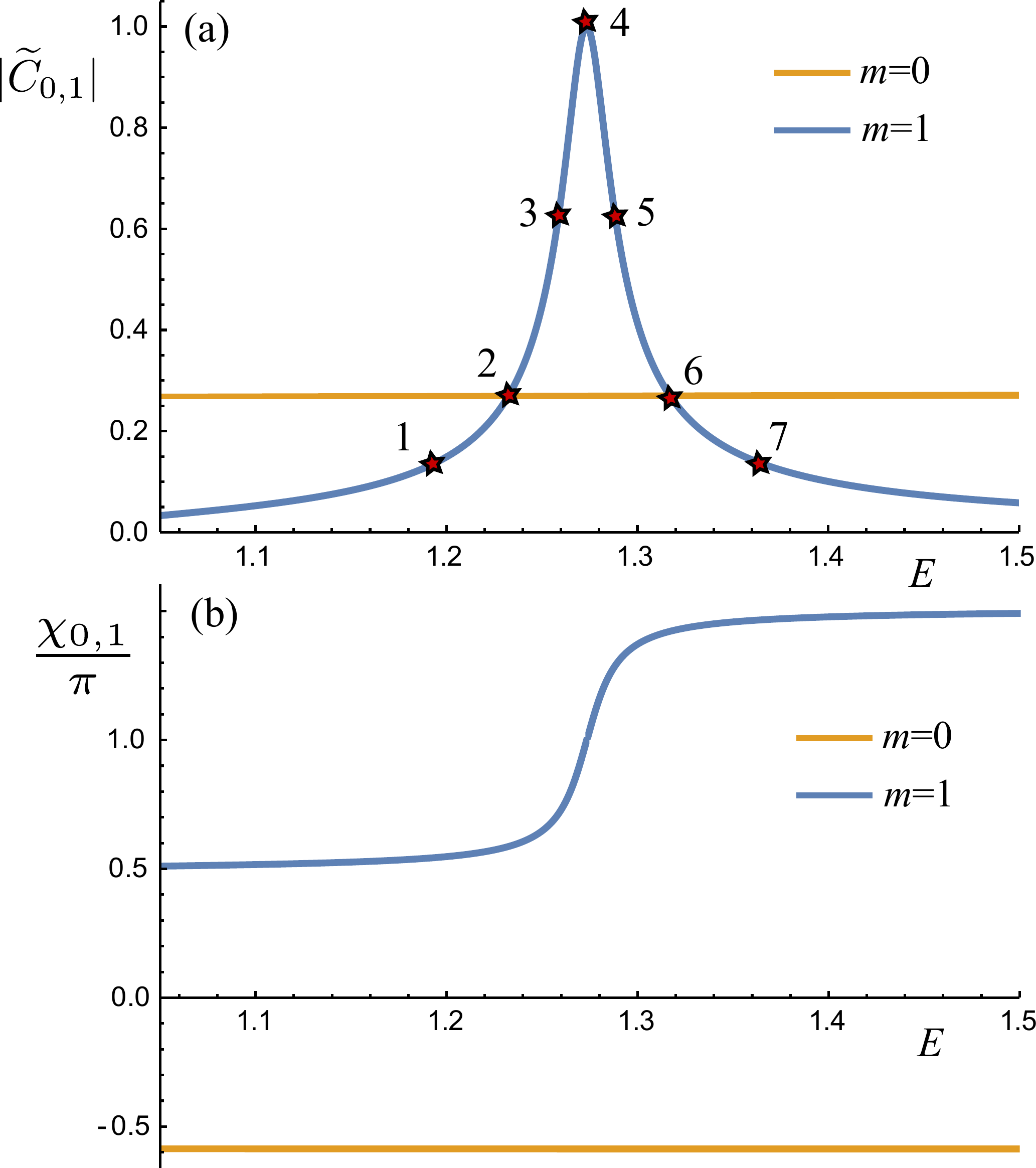}}
    \caption{The amplitude (a) and the phase (b) of the coefficients $\widetilde{C}_1$ and $\widetilde{C}_0$ calculated for parameters $a=0.2$, $v=0.6 \pi$ and $r_0$=0.2. The asterisks marked by numbers 1 --- 7 indicate the energy points at which the angular dependence of $S(E,\varphi)$ changes significantly or its behavior is specific for an energy range.}
    \label{fig3}
\end{figure}

Below we consider these features in more detail using the example of the case with the hybridization parameter $a=0.2$. The key role is played by the function $S(E,\varphi)$ which, in fact, can be considered as a normalized cross section, since $k(E)$ only slowly changes with $E$ in the energy range of interest to us and does not depend on $\varphi$.

The amplitude and the phase of the coefficients $\widetilde{C}_1$ and $\widetilde{C}_0$ are shown in Fig.~\ref{fig3} where the asterisks indicate the points of the energy at which the angular dependence of the scattering cross section significantly changes or is specific to the surrounding energy region. It is seen that the phase of $\widetilde{C}_1$ changes by $\pi$ when the energy passes the resonance, while $\widetilde{C}_0$ is almost constant. 

\begin{figure}[h]
    \centerline{\includegraphics[width=0.95 \linewidth]{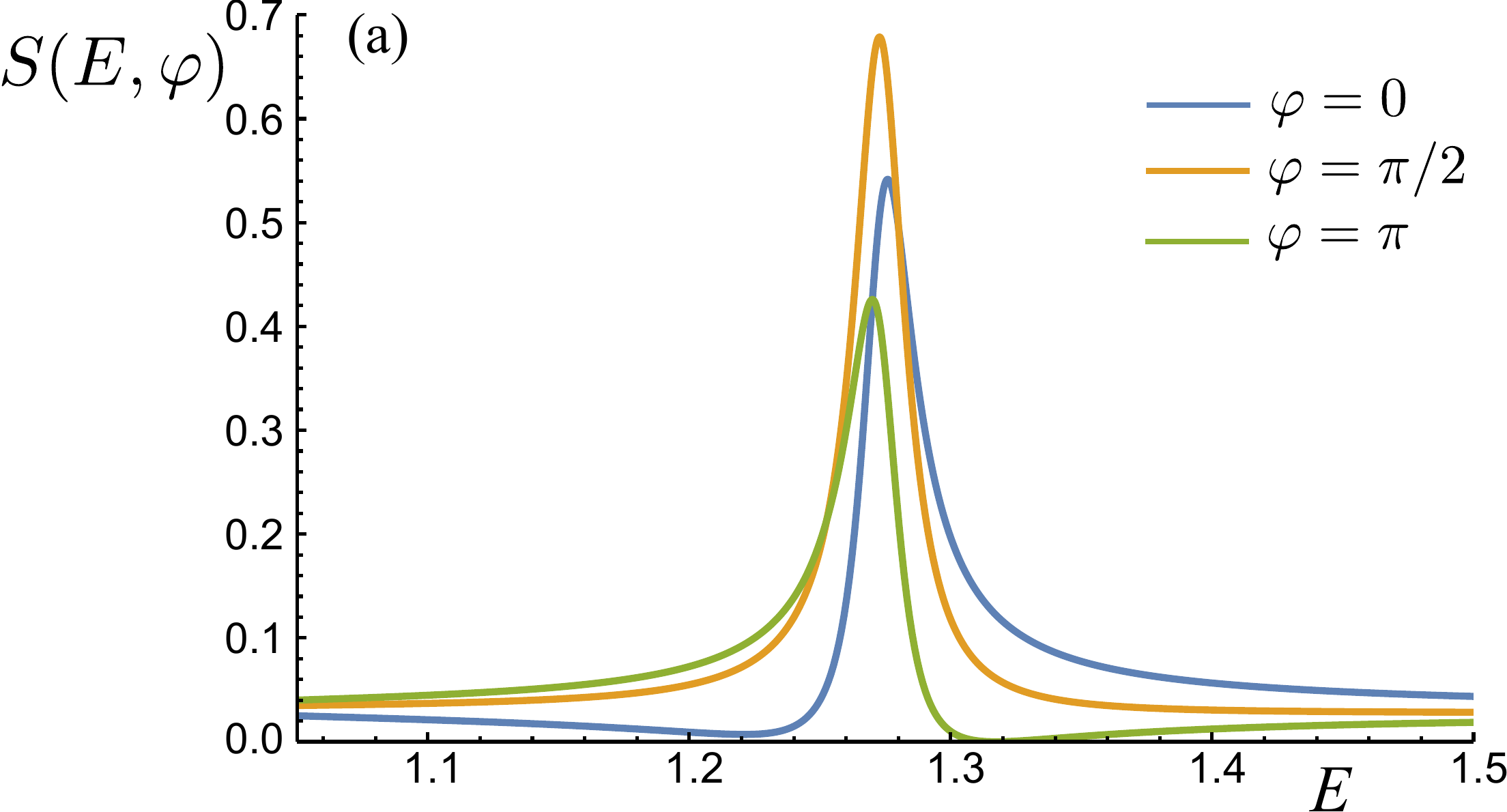}}
    
    \vspace{0.2cm}
    \centerline{\includegraphics[width=0.9 \linewidth]{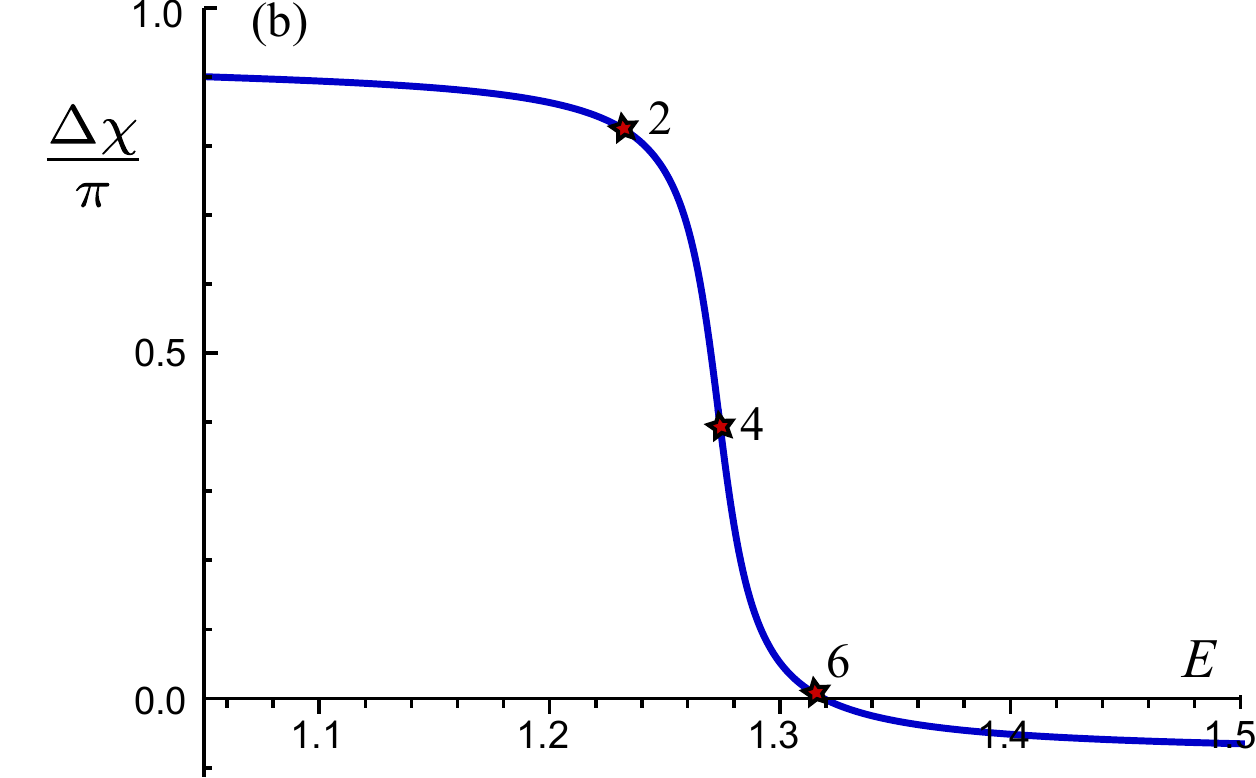}}
    
    \vspace{0.2cm}
    \centerline{\includegraphics[width=0.95 \linewidth]{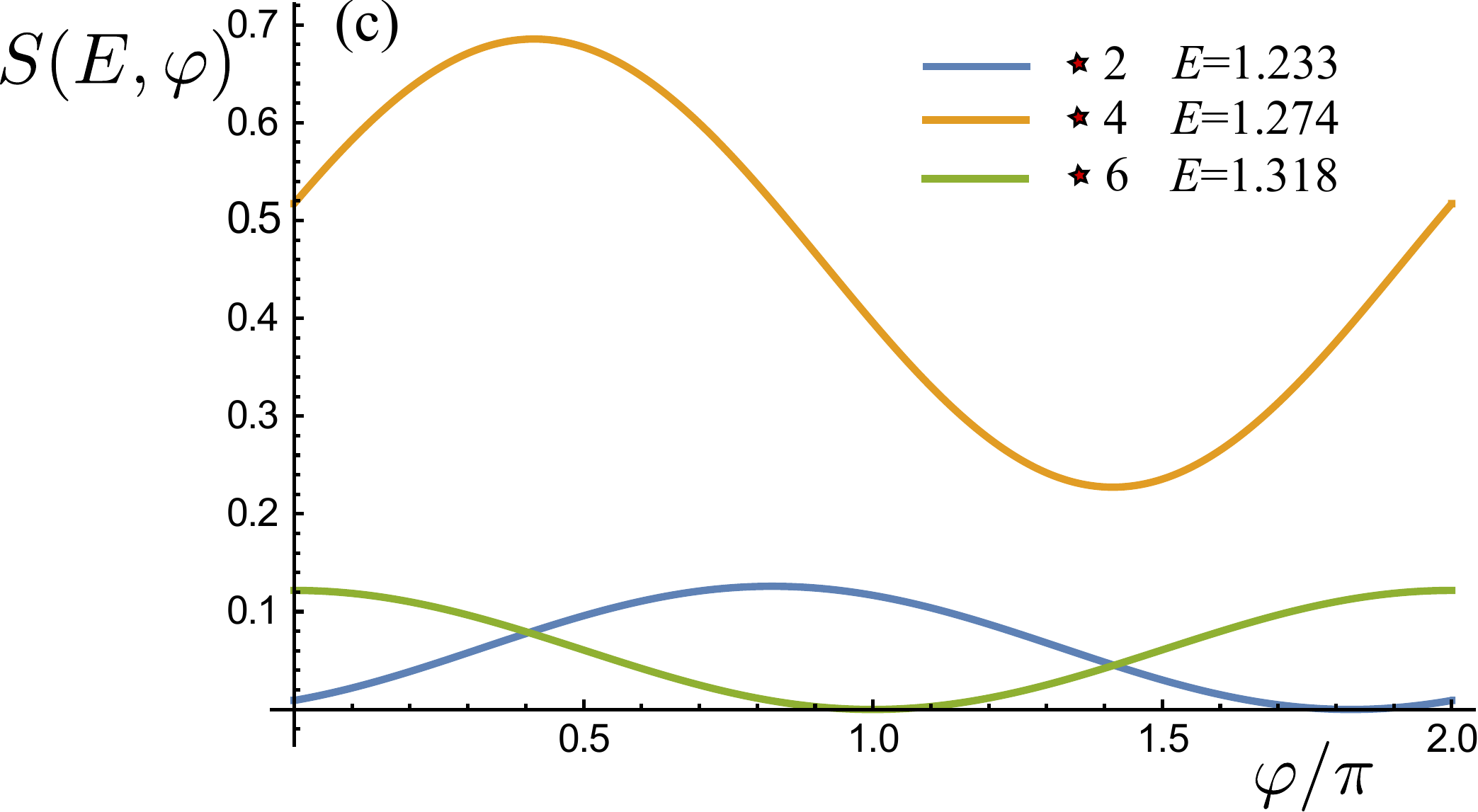}}
    \caption{Energy and angular dependences of the $S(E,\varphi)$ function, which largely determines the scattering cross section. (a) $S(E,\varphi)$ as function of $E$ for three values of the angle. (b) The skewness angle $\Delta \chi(E)$ as a function of $E$. (c) The angular dependence of $S(E,\varphi)$ for three values of the energy near the resonance. The parameters used in the calculation are $a$=0.2, $v=0.6\pi$ and $r_0$=0.2.}
    \label{fig4}
\end{figure}

The main peculiarities of the function $S(E,\varphi)$ are seen from Fig.~\ref{fig4}. The amplitude of $S(E,\varphi)$ as a function of $E$ has a peak that shifts slightly and changes its height as $\varphi$ changes, but the maximum value is reached approximately at $\varphi= \pi/2$. As a function of $\varphi$, at a given energy, the quantity $S(E,\varphi)$ changes in a wide range which indicates a large angular anisotropy. The skewness angle, at which $S(E,\varphi)$ reaches its maximum, changes with energy in the range from 0 to $\pi$. This shows that, depending on $E$, the predominant scattering can occur in the forward, lateral, and backward directions. The angular dependence of $S(E,\varphi)$ is very different for different energies in the resonance region. In the point 4 marked by asterisk in Fig.~\ref{fig3}a, where the coefficient $\widetilde{C}_1(E)$ has the largest value, $S(E,\varphi)$ reaches its maximum at $\varphi \approx \pi/2$. In two other interesting points 2 and 6, where $|\widetilde{C}_1(E)| = |\widetilde{C}_0(E)|$, in contrast, $S(E,\varphi)$ turns to zero at some angle. This is easily to see directly from Eq.~(\ref{e:S}). In these energy points, $S(E,\varphi)=0$ at $\varphi=\Delta \chi(E) \pm \pi$. 

\begin{figure}
    \centerline{\includegraphics[width=0.8 \linewidth]{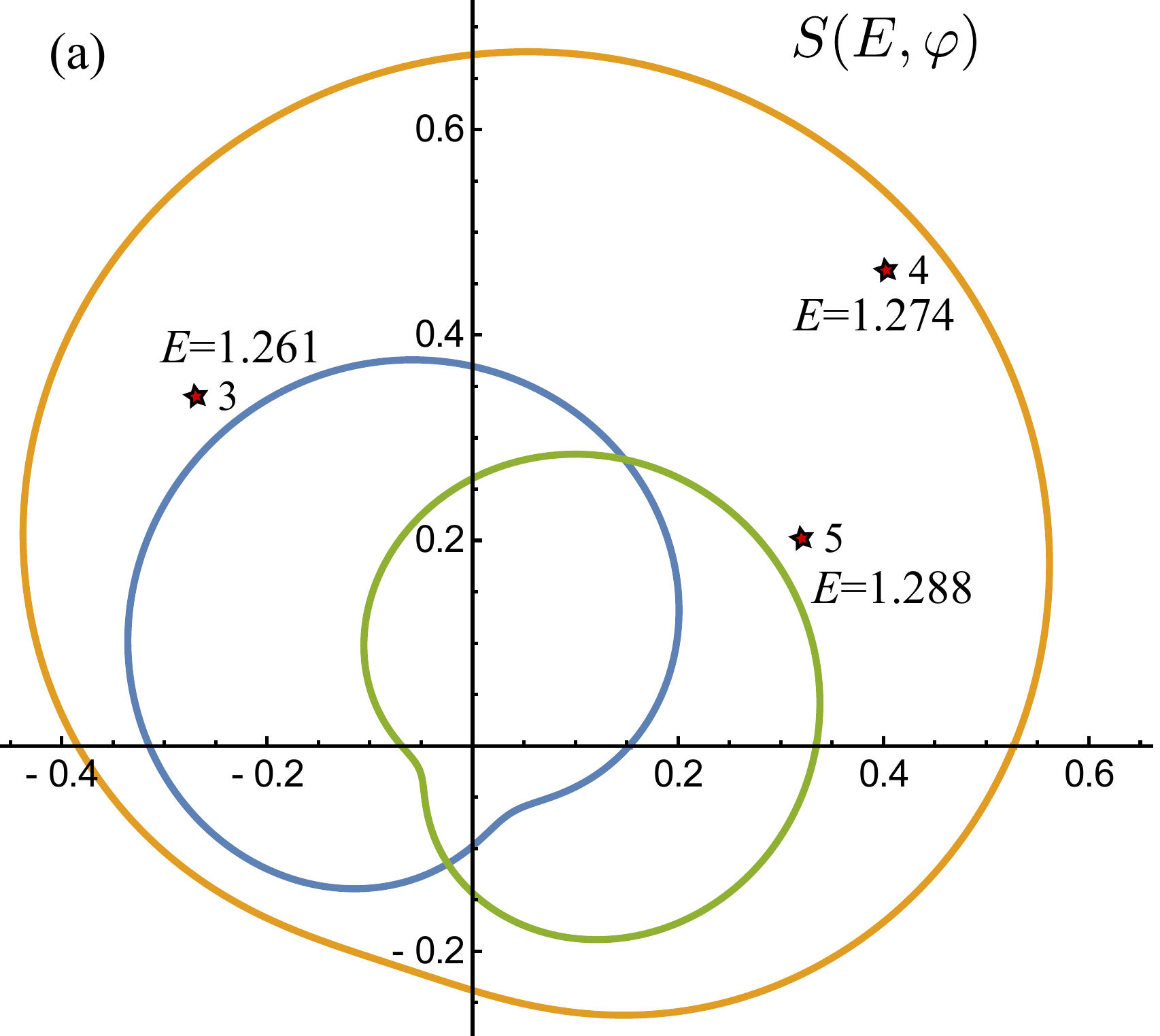}}
    \vspace{0.2cm}
    \centerline{\includegraphics[width=0.8 \linewidth]{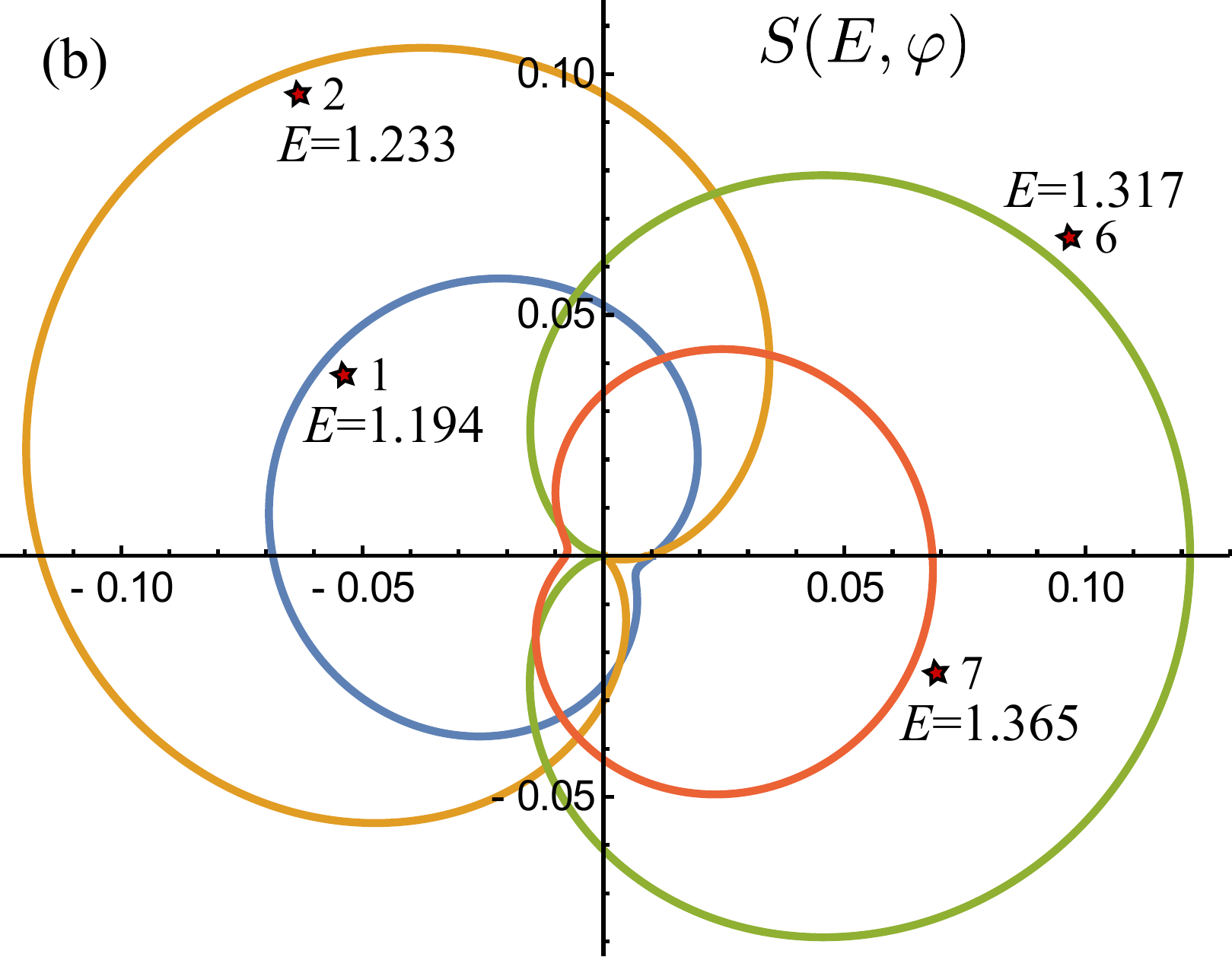}}
    \caption{Polar diagrams of $S(E,\varphi)$ for the energy points marked by asterisks in Fig.~\ref{fig3}. (a) The diagrams for points 3, 4 and 5 near the maximum of the scattering. (b) The diagrams for points 2 and 6 in which the $S(E,\varphi)=0$ at $\varphi=\Delta \chi \pm \pi$, and points 1 and 7 in the region where the scattering becomes weak. The parameters used in the calculation are $a$=0.2, $v=0.6\pi$ and $r_0$=0.2.}
    \label{fig5}
\end{figure}

A more complete picture of the angular dependence of $S(E,\varphi)$ for different energies is given by the polar diagram shown in Fig.~\ref{fig5}. 

It is clearly seen that in the resonance region, Fig.~\ref{fig5}a, the predominant scattering occurs in the direction perpendicular to the incoming flow. But on the slopes of the resonance the situation changes qualitatively, Fig.~\ref{fig5}b. On the low-energy slope, the skewness direction turns into the sector between $\pi/2$ and $\pi$, and the backward scattering becomes predominant. On the high-energy slope, on the contrary, the forward scattering predominates. 

On both slopes, a very nontrivial effect occurs at a certain energy. The scattering cross section vanishes in the direction opposite to the skewness angle. The physical mechanism of this suppression of the scattering is obviously related to the destructive interference of scattered $s$- and $p$-waves, which increases with distance from the peak, and at certain points the scattering cross section in this direction vanishes and then increases again.

For spin-down electrons the polar diagram of the scattering is the same as for spin-up electrons if the sign of $\varphi$ is changed. Thus, in the energy region near resonance, the skew scattering occurs in the opposite lateral direction. But on the slopes, the forward and backward scattering is predominant, just as in the case of spin-up electrons.

\section{Discussion and concluding remarks}\label{S_Conclusion}
We have shown that in materials with a Mexican-hat dispersion, defects with a localized repulsive potential create specific quasi-bound states, which lead to the enhanced skew scattering of electrons with very nontrivial polar diagram and energy dependence. The mechanism of the formation of the quasi-bound states is due to the feature of the Mexican-hat dispersion, which has remained unexplored until now. It consists in the fact that the sign of the effective mass in the region of $\mathbf{k}$-space near the central extremum of the Mexican hat is opposite to the sign of the mass outside this region. Therefore, particles in the states of the region near the central extremum are attracted to the defect, while the particles in the outer region of $\mathbf{k}$-space  are repelled. As a result, a resonant state is formed with an energy above the central maximum of the Mexican hat. 

We have studied these quasi-bound states in the case when the Mexican-hat dispersion arises due to the hybridization of the inverted electron and hole bands, and is described by the BHZ model. The theory of quasi-bound states has been developed by reducing this problem to the Fano--Anderson model in which the hybridization Hamiltonian plays a role similar to a tunneling Hamiltonian. The hybridization Hamiltonian relates the bound state formed by the states of the hole band to the continuum of states of the electron band. Within the framework of this approach, we have found that quasi-bound states are formed mainly by hole-band states with an admixture of electronic states, while continuum states are formed mainly by electron-band states with an addition of hole states. The emerging quasi-bound state creates a resonance of the local density of states, the energy of which is somewhat shifted relative to the energy of the bare bound state in the hole band with an energy above the maximum of the Mexican hat by an amount determined by the hybridization parameter $a$. The width of the resonance is also determined by the parameter $a$ and is changed as $a^2$ when $a^2\ll 1$.

An interesting property of quasi-bound states formed by this mechanism is that they can transform to a BIC\@. However this occurs only for a specific form of the potential at which the resonance width turns to zero.

The most striking manifestations of the quasi-bound states in experiment can be associated with their effect on the electron transport. In this regard, we have studied the electron scattering on a repulsive defect which creates a quasi-bound state. The scattering theory has been developed for a defect with a short-range potential in the limit of a $\delta$-like form. It has been found that a quasi-bound state strongly enhances the skew scattering of electrons with energies near the resonance. But the most non-trivial effect is a strong increase in spin-dependent large-angle scattering with a very unusual polar diagram, which, moreover, varies significantly with energy.

The angular asymmetry of scattering manifests itself in two aspects. One of them refers to spin-dependent asymmetry in the normal direction relative to the incident flow. Electrons with different spins scatter predominantly in opposite directions to the left and right from the incident flow. This is a well known skew scattering effect, but here it is strongly enhanced. The greatest effect is achieved at the resonance energy where the skewness angle is about $\pi/2$. For realistic values of the parameter $a \sim 0.1 - 0.5$, the ratio of scattering cross sections in opposite sides reaches 3, as shown in Fig.~\ref{fig5}. With distance from the maximum, this angular asymmetry decreases, and much more slowly on the low-energy slope than on the high-energy one. Thus, the mechanism under consideration makes it possible to efficiently separate electrons according to their spins.

Another aspect relates to the scattering asymmetry in the direction along the incident flow, that is asymmetry of backward/forward scattering which appears on the slopes of the resonance. Interestingly, on the low-energy slope the backward scattering becomes dominant, while the forward scattering becomes dominant on the high-energy slope as shown in Fig.~\ref{fig5} where the scattering asymmetry in longitudinal direction reaches 3 - 4. 

But the most unexpected result is the suppression of the scattering in the direction opposite to the skewness angle, which occurs due to the interference of scattered $s$- and $p$-waves.

The magnitude of the effect can be estimated from Eq.~(\ref{eq:scatt_cross_section1}) and data of Figs.~\ref{fig3}--\ref{fig5}. The scattering cross section (or more precisely, for 2D systems, the scattering length) is of the order of $k^{-1}$, where $k$ is the wave number for the energy of the order of the central maximum of the Mexican hat. For $a$=0.2 and other parameters close to HgTe, the scattering cross section is estimated as $d\sigma/d\varphi\sim 10^{-6}$~cm. At a defect concentration $N_i\sim 10^{12}$~cm$^{-2}$, this gives an estimate of the mean-free path time for skew scattering of about $10^{-12}$~s, with the asymmetric component of the same order as symmetric one. Thus the considered mechanism of skew-scattering can lead to quite observable transport effects, but this issue requires a further study. The physical mechanism of enhanced asymmetric scattering and the nontrivial features of the angular and energy dependence of the scattering cross section are due to the helical structure of the quasi-bound states in the BHZ model.

\begin{acknowledgments}
This work was carried out in the framework of the state task for the Kotelnikov Institute of Radio Engineering and Electronics.
\end{acknowledgments}
    
\bibliography{mexican-hat_res}

\end{document}